\documentclass{aa}
\usepackage{epsfig,times}
\usepackage{amssymb}
\usepackage{txfonts}
\usepackage{graphicx}
\usepackage{natbib}
\begin{document}

\title{Pulsar timing irregularities and the imprint of magnetic field evolution}

\author{J.A.~Pons\inst{1} \and D. Vigan\`o\inst{1} \and U.~Geppert\inst{2,3}}
\institute{Departament de F\'{\i}sica Aplicada, Universitat d'Alacant,  
Ap. Correus 99, 03080 Alacant, Spain
\and 
German Aerospace Center, Institute for Space Systems, , Robert-Hooke-Str. 7, 28359 Bremen, Germany
\and
Kepler Institute of Astronomy, University of Zielona Gora, Lubuska 2, 65-265 Zielona Gora, Poland}
\date{Received July 24, 2012; Accepted September 4, 2012}

\abstract
{The rotational evolution of isolated neutron stars is dominated by the magnetic field anchored to the solid crust of the star. 
Assuming that the core field evolves on much longer timescales,
the crustal field evolves mainly though Ohmic dissipation and the Hall drift, and 
it may be subject to relatively rapid changes with remarkable effects on the observed timing properties.}
{We investigate whether changes of the magnetic field structure and strength during the star evolution may have
observable consequences in the braking index $n$. This is the most sensitive quantity to reflect
small variations of the timing properties that are caused by magnetic field rearrangements.}
{We performed axisymmetric, long-term simulations of the magneto-thermal evolution of neutron stars with state-of-the-art
microphysical inputs to calculate the evolution of the braking index. 
Relatively rapid magnetic field modifications can be expected only in the crust of neutron stars, where we focus our study. }
{We find that the effect of the magnetic field
evolution on the braking index can be divided into three qualitatively different stages
depending on the age and the internal temperature: a first stage that may be different for standard pulsars 
(with $n\sim 3$) or low field neutron stars that accreted fallback matter 
during the supernova explosion (systematically $n<3$); in a second stage,
the evolution is governed by almost pure Ohmic field decay, and a braking index $n>3$ is expected;
in the third stage, at late times, when the interior temperature has dropped to very low values, 
Hall oscillatory modes in the neutron star crust result in braking indices of high absolute value 
and both positive and negative signs.}
{Current magneto-thermal evolution models predict a 
large contribution to the timing noise and, in particular, to the braking index, from temporal variations of the magnetic field.
Models with strong ($\gtrsim 10^{14}$ G) multipolar or toroidal components, even with a weak ($\sim 10^{12}$ G) dipolar
field are consistent with the observed trend of the timing properties.
}

\titlerunning{Pulsar timing and magnetic field evolution} 
\authorrunning{J.A. Pons, D. Vigan\`o \& U. Geppert} 
 
\maketitle 

\section{Introduction.}

The spin-down of an isolated neutron star (NS) on secular timescales is mainly caused by rotational energy losses due to electromagnetic radiation, pulsar winds, or
gravitational radiation. 
A measurable quantity closely related to the rotational evolution of pulsars is the braking index $n$,
defined by assuming that the star spins down according to a power law $\dot{\Omega}= -K \Omega^n$,
where $\Omega$ is the  NS angular velocity. 
For each of these dominant rotational energy loss mechanisms, the proportionality 
constant $K$ hides different dependences on the star radius, moment of inertia, magnetic field strength, 
and angle between rotation and magnetic axis.
If all these quantities are constant in time, the magneto-dipole spindown mechanism 
predicts a braking index $n=3$, but variations in time of any of these quantities may cause 
departures from this canonical value.  
Unfortunately, the accurate determination of the second derivative of the frequency, needed to estimate the  braking
index, is not always possible because it is affected by glitches and other short-term timing irregularities. 

At present, eight pulsars have sufficiently steady rotations that stable values of their braking index are generally accepted \citep{Lyne93,Lyne96,Mid2006,Liv2007,Liv2011,Welt2011,Espinoza11}, among which the most recent case is
PSR J1734-3333, which has $n=0.9\pm0.2$ significantly below 3 \citep{Espinoza11}.
All these cases show the same trend: they are all young pulsars (Vela is the oldest among these pulsars 
with $t\approx 12$ kyr) and they all have $n<3$.
We note that gravitational wave emission predicts $n=5$, but
it is only efficient during the first minutes or hours of a NS life, when rotation 
is sufficiently fast and the mass quadrupole moment large enough (see e.g., \cite{HJA06} and references therein). 
It certainly does not contribute significantly to the timing noise of pulsars older than $100$ yr.

Selecting a sample of 127 pulsars from the ATNF Pulsar Catalogue \citep{ATNF},  
for which the quoted errors in  the second derivative of the spin frequency ($\ddot{\nu}$) are less than ten per cent,
\cite{Urama2006}
found a strong correlation of $\ddot{\nu}$ with $\dot{\nu}$, independent of the sign of $\ddot{\nu}$. They suggested
that this trend can be accounted for by small stochastic deviations in the spin-down torque that are directly
proportional (in magnitude) to the spin-down torque itself.
Another point discussed in the literature is that some of the old pulsars ($>10^6$ yr) have braking indices with absolute values exceeding $|n|=10^4$.
The occurrence of very high braking indices of both signs has been considered in the context of internal frictional instabilities occurring between the crust and the superfluid, almost independently of the evolution of the neutron star magnetic field \citep{SM95}. However, this applies only for old neutron stars ($\tau \gtrsim 2 \times 10^7$ yr) and appears as extremely short term events oscillating about the canonical value $n=3$.
\cite{BT10} proposed another explanation for the observed distribution with very high positive and negative braking 
indices by studying the effect of nondipolar magnetic field components and neutron star precession
on magnetospheric electric current losses. 
These large $n$ should be observable over relatively long periods of $10^3$-$10^4$ yr.
Another possibility that can explain the observed variability of braking indices is
the time-evolution of conductivity in the magnetosphere \citep{Spit2012}, which also has
implications for the spin-down of intermittent pulsars and subpulse drift phenomena \citep{Lyne2009}.

In this paper we focus on the imprint that the time-evolution of the internal magnetic field has 
on the timing noise. Our goal is to estimate the contribution of this variability (expected in realistic evolutionary models)
to the braking index at different epochs. In particular, we discuss the possible reemergence of a magnetic field 
initially submerged by hypercritical accretion during the supernova explosion, and the evolution of the crustal
magnetic field under the combined action of Ohmic dissipation and Hall drift during the first few million years of a NS life.

\section{Basic equations.}

The spin--down behavior of a rotating neutron star is governed by the energy balance equation 
relating the loss of rotational energy,  ${E}_{\rm{rot}} = I\Omega^2/2$,
where $I$ is the moment of inertia of the neutron star and $\Omega$ the angular velocity, to the 
energy loss rate $ \dot{E}$, given by magneto-dipole radiation, wind, gravitational radiation, or others:
\begin{equation}
\dot{E}_{\rm{rot}} = I \Omega \dot{\Omega} \approx \dot{E}.
\hspace*{-5mm}\label{equ:spin-down}
\end{equation}

The standard way to define the braking index is 
\begin{equation} 
 n = \frac{\ddot{\Omega}\Omega}{\dot{\Omega}^2} = \frac{\ddot{\nu} \nu}{\dot{\nu}^2} =
 2 - \frac{\ddot{P}P}{\dot{P}^2} \;,
\hspace*{-5mm}\label{equ:BIstandard}
\end{equation} 
where  $\nu$ is the spin frequency, and we denote by $P=1/\nu$ the period.
Under the usual assumption of a power-law rotational evolution
\begin{equation}
\dot{\Omega} = - K \Omega^{n}\,,
\label{equ:omegadot-hat}
\end{equation}
the power-law index coincides with the braking index.

For the particular case of a rotating dipole in vacuum, the well--known Larmor formula returns
\begin{equation}
\dot{E}= \frac{B_0^2R^6\Omega^4}{3c^3} \sin{\alpha}^2\;,
\label{equ:mdr}
\end{equation}
where $B_0$ is the dipolar component of the surface magnetic field at the pole,  $R$ denotes the neutron star radius,
$\alpha$ is the angle between the rotational and the magnetic axis, and $c$ is the speed of light. 
We neglect the contribution to the spindown of higher order multipoles,
since they have significantly shorter ``lever arms" than the dipolar mode.

Combining Eqs. (\ref{equ:spin-down}) and (\ref{equ:mdr}), one obtains
\begin{equation} 
\dot{\Omega} = - K \Omega^3  
\label{equ:omegadot}\;,
\end{equation}
where $K=f_\alpha B_0^2R^6/3Ic^3$.  For simplicity, but without loss of generality, we have omitted the
geometric factor that carries the dependence on the inclination angle, and we
assume hereafter that it is a correction $f_\alpha$ factorized in $K$, which carries information about the
particular physical process that governs energy losses.
The functional dependence of $K$ for all other magnetic processes is the same $\propto B_0^2\Omega^4R^6$. 
Differences in the radiation mechanism are included in the factor $f_\alpha$. While 
magnetospheric current losses scale as $\cos^2\alpha$ \citep{BN07}, 
magneto-dipole radiation losses scale as $\sin^2\alpha$ for vacuum or 
$\frac{3}{2}(1+\sin^2\alpha)$ for force-free magnetospheres \citep{Spitko2006}.
The most recent resistive solutions for pulsar magnetospheres \citep{Spit2012} fit the spin-down luminosity with a prefactor
of the order of unity that also depends on the maximum potential drop along field lines in the corotating frame.

Eq.~(\ref{equ:omegadot}) can also be cast in the usual form
\begin{equation}
P \dot{P} = {\cal{K}} B_0^2(t)
\label{equ:spin-down-B(t)} 
\end{equation} 
where for a standard neutron star ($R=10^6$cm, $I=10^{45}$gcm$^2$) the constant 
${\cal{K}} \approx 10^{-39}$ cm s$^3$g$^{-1}$. Here we have explicitly written $B_0(t)$ to show that the 
magnetic field is the only quantity that we allow to vary with time.

Now we consider what happens when a secular variation of the magnetic field is allowed. Deriving $\ddot{P}$ from
Eq.~(\ref{equ:spin-down-B(t)}), we have
\begin{equation} 
\ddot{P}={\cal{K}}BP^{-2}\left(2\dot{B_0}P\,\,-\,\,B_0\dot{P}\right)~,
\end{equation} 
and the braking index in a general case of a time-dependent magnetic field can be simply expressed as follows
\begin{equation}
n=3-4 \frac{\dot{B_0}}{B_0} \tau_c \equiv 3-4 \frac{\tau_c}{\tau_B}~,
\hspace*{-5mm}\label{equ:hatn} 
\end{equation}
where $\tau_c = P/2 \dot P$ is the characteristic age, and we defined 
the magnetic field evolution timescale 
$\tau_B \equiv B_0/\dot{B_0}$.
This last equation shows that any 
variation of the dipolar surface magnetic field strength results in a deviation from the $n=3$ 
standard value,  which is obviously recovered for a constant magnetic field ($\dot{B_0} = 0$). 
For an increasing $B_0$ we will always obtain ${n} < 3$, while ${n} > 3$ is the signature 
of a decreasing $B_0$. 

\section{Magnetic field evolution scenarios.}

The precise calculation of the time variation of $B_0$ requires solving the problem of the coupled magneto-thermal
evolution of a neutron star \citep{PMG2009,VPM2012}. 
However, before showing results from simulations, one can qualitatively analyze
of the most important physical process at different ages. 
We now discuss three possible scenarios.

\subsection{Amplification or reemergence of the dipolar surface magnetic field.}

The generic $n< 3$ observed without exception for young pulsars can be
caused by the rediffusion of the magnetic field submerged in the crust during the supernova fallback episode
\citep{young95,muslimov95,GPZ99}. The submergence of the magnetic field has also been studied
in recent MHD simulations, with interesting implications for gravitational wave emission \citep{Vigelius2009,Vigelius2010}.
The rediffusion timescale depends essentially on the total amount of accreted matter. 
For total accreted masses between $10^{-4}$-$10^{-3} M_\odot$, screening currents are dissipated on 
$\tau_{\rm rediff} \sim 10^3$-$10^4$ yr, as recent calculations have confirmed \citep{Ho2011,Vigano12}.
The same idea has been applied to the reemergence of magnetic fields in accreting binary 
systems with similar results \citep{CZB2001}, and evidence for
fast magnetic field evolution in an accreting millisecond pulsar has been reported \citep{Patruno2012}.

Alternative mechanisms are time variations in the angle between the rotational and magnetic axes
\citep{LE1997,RZC1998},
or the thermoelectric field generation that may proceed in the crust and envelope 
of young pulsars if a sufficiently strong temperature gradient is present  \citep{ULY86,WG96}.
This process  is limited by the condition that the surface temperature of the neutron star should not be lower than 
$3\times 10^6$ K \citep{WG96}, which corresponds to about $1000$ yr in the standard cooling scenario.
In our simulations we included the reemergence of a screened magnetic field, but our present version of the
code does not include magnetic field generation by thermoelectric effect, or other possible mechanisms
such as magnetic flux expulsion from the superconducting core.

\subsection{Ohmic decay.}

Ohmic dissipation of the magnetic field has been thoroughly investigated in the past
by many groups \citep{HUY1990,PGZ00,Tau2001}. 
The usual definition of the Ohmic decay time is
\begin{equation}
\tau_{\rm{Ohm}}= \frac{4\pi\sigma L^2}{c^2}\;,
\hspace*{-5mm}\label{equ:tauohm}
\end{equation}
where $L$ denotes the scale-length of magnetic field variations and $\sigma$ the electric conductivity.
In a neutron star crust, $\sigma$ is dominated by electronic transport and depends on the 
electron density, the crustal temperature, and the impurity concentration within the crust. 
Since the electron density varies over about four orders of magnitude in the crust and the temperature decreases by about two to three orders of magnitude in a pulsar's lifetime, the electric conductivity may vary both in space and time by many orders of magnitude (see Fig. 1 in \cite{PG07}). Therefore, to assume a uniform Ohmic decay time $\tau_{\rm{Ohm}}$, independent of the location of currents and the pulsar age is, generally speaking, quite misleading:
Ohmic decay in a neutron star crust cannot be described by a single exponential law 
(see e.g. \cite{PGZ00} for a qualitative discussion).
Since the conductivity increases with decreasing temperature,  Ohmic diffusion becomes 
an increasingly slow process in older pulsars, beyond the transition from neutrino dominated 
to photon cooling at $\gtrsim 10^5$ yr. The only general fact that we can expect is that,
for low-field middle-aged neutron stars ($10^4$-$10^5$ yr), 
this is expected to be the dominant process, thus resulting in $n>3$.

\subsection{Hall drift oscillatory modes.}
The Hall drift, through its nonlinear dependence on the magnetic field,
has an influence on the magnetic field evolution either for magnetar conditions ($B>10^{14}$ G), or alternatively when the
electrical resistivity becomes very low.
Under certain circumstances the Hall drift can drain magnetic energy out of the dipolar mode and redistribute it into smaller scale ones. This sometimes causes an apparently rapid decrease of $B_0$. But the 
opposite effect may also occur, i.e., pumping energy from an internal strong toroidal field to the dipolar poloidal component.
Both scenarios would happen on the Hall timescale
\begin{equation}
\tau_{\rm{Hall}}= \frac{4\pi n_e e L^2}{cB}\;,
\hspace*{-5mm}\label{equ:tauHall}
\end{equation}
where $n_e$ is the electron density and $e$ the elementary charge. For typical values in NSs we have
$\tau_{\rm Hall} \sim (10^4$-$10^6) \frac{10^{14}{\rm G}}{B_0}$ yr.

In situations of quasi-equilibrium, oscillatory modes with magnetic energy exchange between the 
different crustal field modes are expected. The temporal evolution of the polar surface magnetic field can then be
approximated by
\begin{equation}
B_0(t)= B_0 + \delta B\sin \left( \frac{2\pi t}{\tau_{\rm Hall}}\right)\;,
\hspace*{-5mm}\label{equ:tauHall2}
\end{equation}
and from Eq.~(\ref{equ:hatn}) we have
\begin{equation}
n=3 -4\frac{\dot{B_0}}{B_0} \tau_c \approx 3-8\pi \frac{\tau_c}{\tau_{\rm{Hall}}} \frac{\delta B}{B_0} \cos\left( \frac{2\pi t}{\tau_{\rm Hall}}\right)\;.
\hspace*{-5mm}\label{equ:tauHall3}
\end{equation}

These oscillations are expected in young magnetars with $B_0>10^{14}$ G. It causes 
corrections to the canonical braking index $n=3$ of either positive or negative sign, according to Eq.~(\ref{equ:tauHall3}).
These corrections are
expected to be small because of their short $\tau_c$, and to be increasingly
important for objects with smaller $\dot{P}$ (old characteristic ages).
In addition, as the star cools down, the drop in electrical resistivity may activate the Hall term 
even for normal pulsars $B_0 \gtrsim 10^{12}$ and the occurrence of oscillatory modes 
during a {\it second Hall stage} at late times is a natural outcome. 
First estimates suggest that $\delta B/B_0$ could be as large as $\sim 10^{-1}$. 
When $\tau_c \gg \tau_{\rm Hall}$, 
the second term in Eq.~(\ref{equ:tauHall3}) dominates and the magnetic field oscillatory modes should
result in equally probable positive and negative braking indices with high absolute values. However,
this needs to be confirmed by long term realistic simulations.

\begin{figure}[th]
\centering
\includegraphics[width=0.4\textwidth]{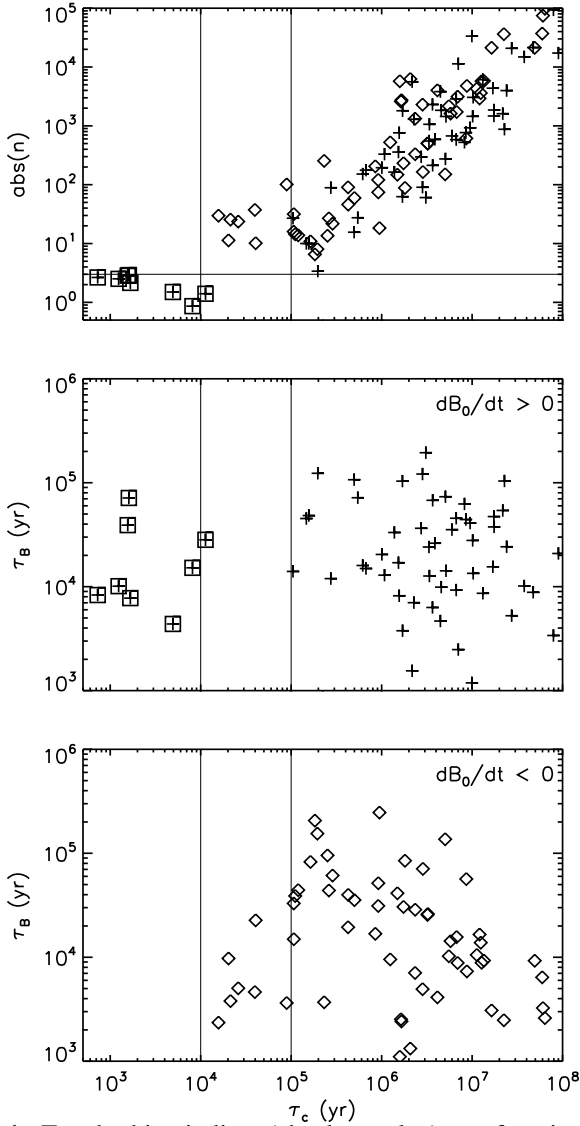}
\caption{
Top: braking indices (absolute value) as a function of $\tau_c$ for our sample of 118 pulsars. 
We represent pulsars with $n<3$ with crosses and pulsars with $n>3$ with diamonds. Note that 
all objects marked with crosses and $\tau_c>10^5$ yr have actually $n<0$.
The 8 youngest objects discussed in \cite{Espinoza11} are marked with squares. 
Middle:  magnetic field evolution timescale, $\tau_B=4 \tau_c/(3-n)$ for pulsars with $n<3$. 
Bottom: Absolute value of $\tau_B$ (they are all negative) for pulsars with $n>3$.  }
\label{fig1}
\end{figure}

\section{Discussion.}

\subsection{Magnetic field evolution in the pulsar population?}

From the whole population in the ATNF Pulsar Catalogue \citep{ATNF}, we extracted a sample of
pulsars for which $\ddot{\nu}$ has a quoted error smaller than 10\%. From this preselected sample, 
we excluded all pulsars in binaries and those with very short periods ($P < 15$ ms), likely to be recycled, 
and we doubled checked our list with the more recent review by \cite{Hobbs10}, the first large-scale analysis
of pulsar timing noise over timescales $>10$ yr, which led us to remove several more pulsars whose revised values were inconsistent
with the ATNF data, or with larger errors. After this selection, our sample contains 118 radio-pulsars, 
about half of which have negative braking indices.

We show in the top panel of Fig.~\ref{fig1} the observed distribution of braking indices as a function of $\tau_c$. We  considered the characteristic magnetic field evolution timescale, $\tau_B= 4 \tau_c/(3-n)$ (see Eq.~(\ref{equ:hatn})) and 
separated the sample into two groups, those with positive (middle panel) and negative (bottom panel) $\tau_B$. The eight pulsars discussed in \cite{Espinoza11} are marked with squares. 
With all due cautions regarding the uncertainties associated to these measurements, it is worth mentioning some interesting trends visible in this plot:

\begin{itemize}
\item
All young objects seem to have always $n<3$, which can be a hint of an increasing dipolar magnetic field.
\footnote{We use the values collected in \cite{Espinoza11} for this plot.}
\item
All middle age objects ($10^4$-$10^5$ yr), except Vela, that survived to our conservative selection
criteria have negative time derivatives of $B_0$.\footnote{We did not include seven more objects, 
that fall in this region, with a quoted error of $\ddot{\nu}$ 
smaller than 10\% in the ATNF pulsar database. They are not considered in \cite{Hobbs10}, probably because they did not have 10 yr of accumulated data, although their values have not been reported to change.}
\item
For old objects, there is no correlation at all, and there are similar numbers of objects with positive and negative 
derivatives of the field. The typical evolution timescales are in the range $10^3$-$10^5$ yr.
\end{itemize}

The strong correlation between $|n|$ and $\tau_c$ seen in the old objects of the top panel simply reflectis the definition of $n$, Eq.~(\ref{equ:hatn}), with the additional piece of information that $\tau_B$ does not seem to be
correlated with $\tau_c$ (see middle and bottom panels).

Note that our criteria automatically select the objects with a dominant contribution of $\ddot{\nu}$ in the timing phase residuals (i.e., the cleanest cubic lines in Fig.~3 of \cite{Hobbs10}, e.g. B0114+58). This introduces a bias toward objects with high braking indices, and against pulsars with residuals dominated by higher order terms (e.g. B0136+57) or quasi-periodic terms (e.g. B1642-03 or B1826-17) in the time-dependent phase \citep{Lyne2010}. In the latter cases, the estimates of $\ddot{\nu}$ are subject to larger uncertainties, and $n$ strongly depends on the analyzed time interval \citep{Hobbs10}. However, we have checked that the general trends do not change if the sample is enlarged by including pulsars with quoted errors in $\ddot{\nu}$ up to 80\%. This simply increases the statistics ($>300$ sources) and includes some objects with lower value of $|n-3|$ (shorter $|\tau_B|$).

\subsection{Expected evolution from theoretical models.}

We performed a series of numerical simulations with the magneto-thermal evolution code presented in \cite{Aguilera2008b,PMG2009}, and \cite{VPM2012}, which includes all relevant microphysical processes and follows the complex feedback between the physical mechanisms presented in a simplified way in the previous section. In addition to Ohmic dissipation and Hall drift effects, which are consistently included in the simulations, we assumed that all models suffer an episode of hypercritical accretion during their formation.

\begin{figure}[th]
\centering
\includegraphics[width=0.45\textwidth]{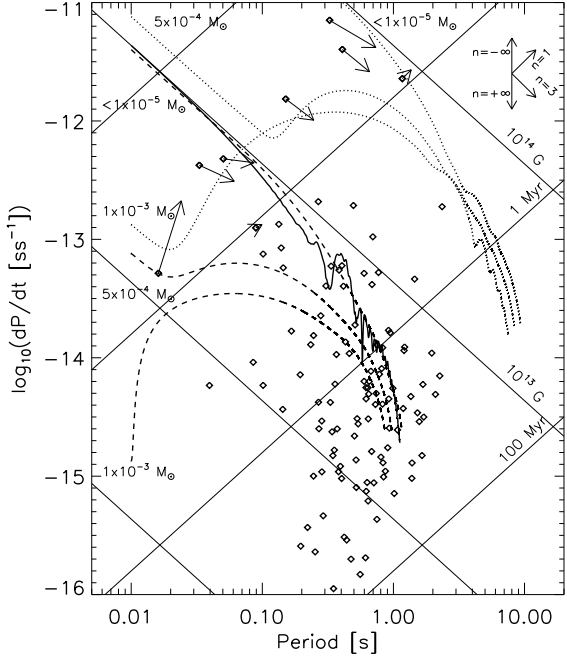}
\caption{Evolutionary tracks in the $P$-$\dot{P}$ diagram during the first 3 Myr for: model A (dashed) and model B (dotted), both for different values of 
accreted mass $M_a=[0.1,5,10] \times 10^{-4} M_\odot$; model C (solid) without accretion.}
\label{fig2}
\end{figure}

We plot in Fig.~\ref{fig2} the evolutionary tracks of some theoretical models in the $P$-$\dot{P}$ diagram, compared to the pulsars of our sample. 
We show results for three initial magnetic field configurations. Two of them are purely dipolar, although during the evolution other multipoles and toroidal field are naturally created. They differ by the magnetic field strengths:
$B_0=10^{13}$ G (model A, dashed lines) and $10^{14}$ G (model B, dotted lines). In each case (models A and B)
we compare results with three values of the total accreted mass. 
The third initial configuration (model C, solid line) has an initial dipolar field of $B_0=10^{13}$ G, as model A, 
but with an additional octupolar component of strength $3.5\times10^{14}$ G (at the pole).
This model serves to make explicit the effect of the complex initial geometries.
Assuming vacuum external boundary conditions, the 
toroidal field has to vanish at the star surface, but the dipolar component
is still coupled to higher order multipoles, which can also interchange energy with the dipolar mode.

\begin{figure}[t]
\centering
\includegraphics[width=0.45\textwidth]{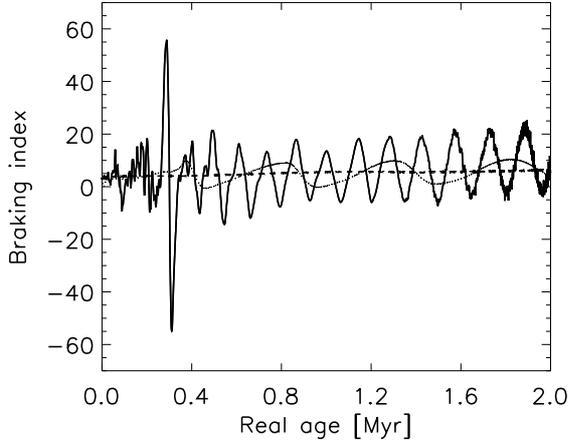}
\caption{Braking index ($n$) as a function of age for model A (dashed line), model B with $M_a = 10^{-3} M_\odot$ (thin 
solid line), and model C (thick solid line).} 
\label{fig3}
\end{figure}

For the eight youngest pulsars, we indicate their predicted movement for the next 2.5 kyr with arrows, assuming the present value of $n$, from \cite{Espinoza11}, remains constant. The direction of the tangent vector to a given track is related to the braking index. For reference, we indicate in the legend on the upper right corner that $n=3$ and $n=1$ imply constant inferred $B_0$, and constant characteristic age, respectively.

The initial period $P_{in}$, which was assumed to be 0.01 s in all cases, only affects the early stage, while $\dot{P} \lesssim P_{in}/t$ (with $t$ being the real age), and $P\simeq P_{in}$. Models with deep submergence of the field
but with different initial periods also have vertical trajectories shifted to the left/right depending on $P_{in}$, and 
quickly cross the range of $\dot{P}$ where the bulk of pulsars lie. 
At late times, tracks coming from the same model but with different $P_{in}$ are indistinguishable, 
typically converging after $t \sim 30 (\frac{P_{in}}{0.01 s}\frac{10^{13}G}{B_{in}})^2$ yr, with $B_{in}$ being
the initial magnetic field.
Therefore, $P_{in}$ has an appreciable long-term effect only in the deep submergence case ($M_a \gtrsim 10^{-3} M_\odot$), for which $B_{in}$ is strongly reduced: in these models, the first few $10^4$ yr are spent in the vertical trajectories, 
with $n\ll 0$. Note that in this reemergence phase there is no correlation between the real and characteristic ages.

When reemergence of the field has almost been completed, the trajectories reach the high-$\dot{P}$ region (i.e., largest $B_0$) and progressively bend. The more extreme braking indices of PSR J1734-3333 ($n=0.9\pm0.2$, \cite{Espinoza11}) and PSR J0537-6910 ($n\sim -1.5$, \cite{Mid2006}) would be consistent with the last stage of the reemergence after a deep submergence into the inner crust.
On the other hand, in the shallow submergence models ($M_a \sim 10^{-5}$-$10^{-4} M_\odot$),  
after accretion stage $B_0\simeq B_{in}$. These tracks initially run almost along the iso-magnetic lines: pulsars with $n$ slightly less than 3 are compatible with this scenario.

Independently of the early reemergence phase (if any), tracks with the same $B_{in}$ converge at middle-age, and
have slopes corresponding to $n>3$,  characteristic of the slow Ohmic dissipation. During this phase, there is a correlation between real and characteristic ages, with typically $\tau_c$ being a factor of few longer than the real age. Some tracks show visible oscillations produced by the Hall activity when the star is cold enough ($t\gtrsim 10^5$ yr); in particular, model C (solid line) clearly shows that any complex initial geometry may have a distinct 
signature on the timing properties of pulsars.

We emphasize that we did not attempt to fit individual objects: our purpose with this sample of models is simply to show that, with reasonable assumptions, it is possible to explain the variability in the observed range of braking indices of young pulsars
and to predict their evolutionary paths in the $P$-$\dot{P}$ diagram.

\subsection{Braking index and evolution timescale for realistic magnetic field evolution models.}

In Fig.~\ref{fig3} we plot the braking index evolution for three representative models,
Note that the horizontal axis in Fig.~\ref{fig1} represents the characteristic age $\tau_c$, while in Fig.~\ref{fig3} we show our results as a function of the real age of each model, so that a direct comparison is not possible.
We also used a linear scale in this plot to show the quasi-periodic oscillations during the long-term evolution more clearly.
For fields $\lesssim 10^{13}$ G and simple dipolar geometries (dashed line) the braking index at late times is
$n>3$, but its absolute value is low. In contrast, for strong dipolar fields, or for weak dipolar components but
with strong higher order multipoles, Hall-drift induced
oscillations appear sooner or later and,  in some situations, have large amplitudes that result in very high absolute values
of the braking index. It is particularly interesting to compare models A and C (dashed and solid line),  which have the same 
initial dipolar component, representative of a typical pulsar (at $10^6$ yr the dipolar field is about $3$-$5\times 10^{12}$ G).
However, the presence of a strong octupolar component at birth results in a radically different braking index behaviour, 
even if $P$ and $\dot{P}$ are similar.
The amplitude of the oscillations and whether the modes are damped or excited depends on particular details of the
small-scale structure of the magnetic field, which is unknown. The exact age at which these oscillatory modes are excited
is connected to the temperature of the star, and therefore to its cooling history and internal physics (neutrino emission 
processes, superfluid gaps, etc.). For the standard cooling scenario this happens at $\approx 10^5$ yr. We can also 
observe that as the star evolves and the magnetic diffusivity decreases, the frequency of the dominant
mode may vary (see solid line). For example, at early times in model C we observe
variability on shorter timescales (few kyr), while later we see longer oscillation periods ($10^5$ yr). After 1.5 Myr
higher frequency modes seem to be growing again.

\begin{figure}[th]
\centering
\includegraphics[width=0.45\textwidth]{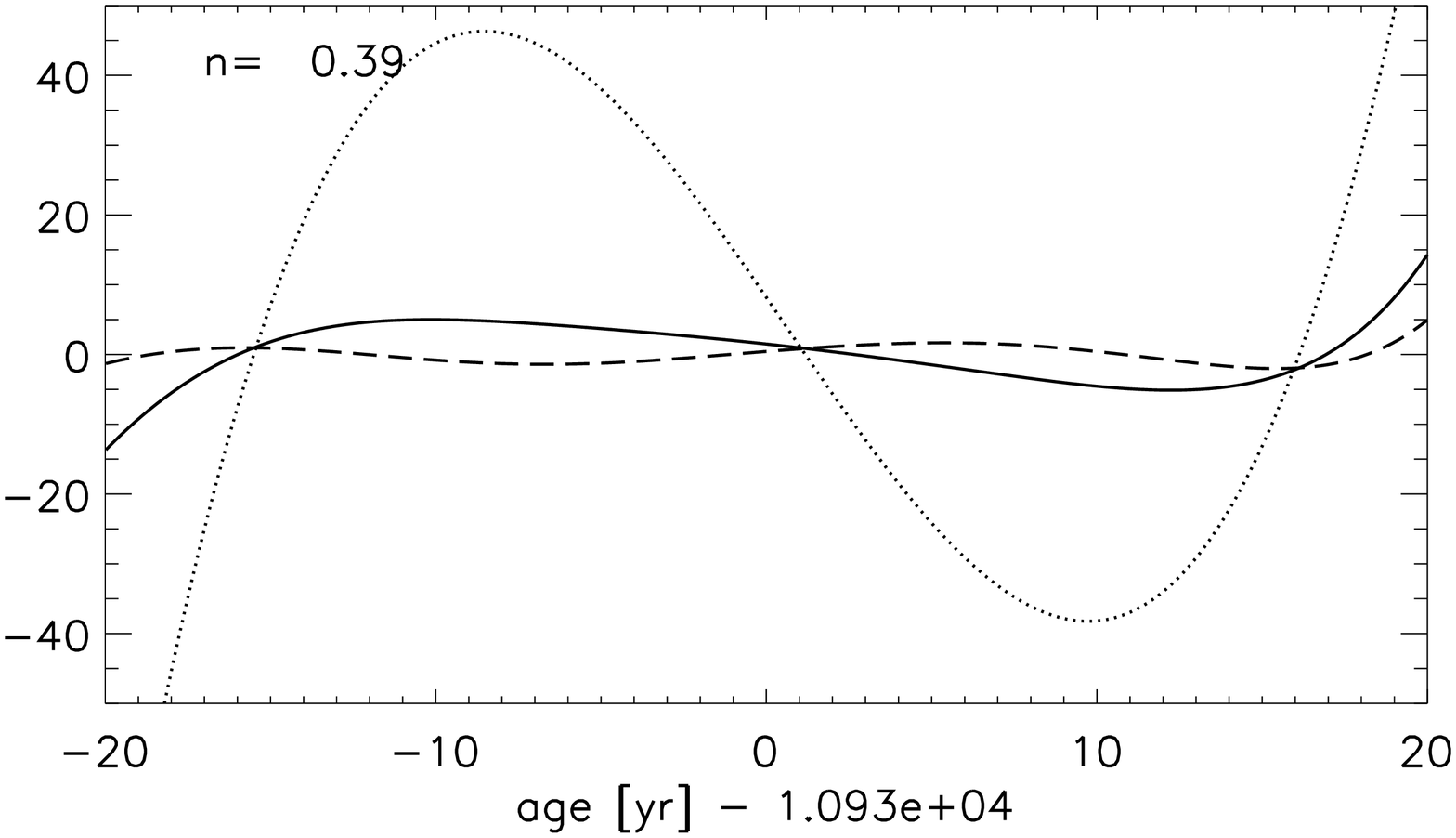}
\includegraphics[width=0.45\textwidth]{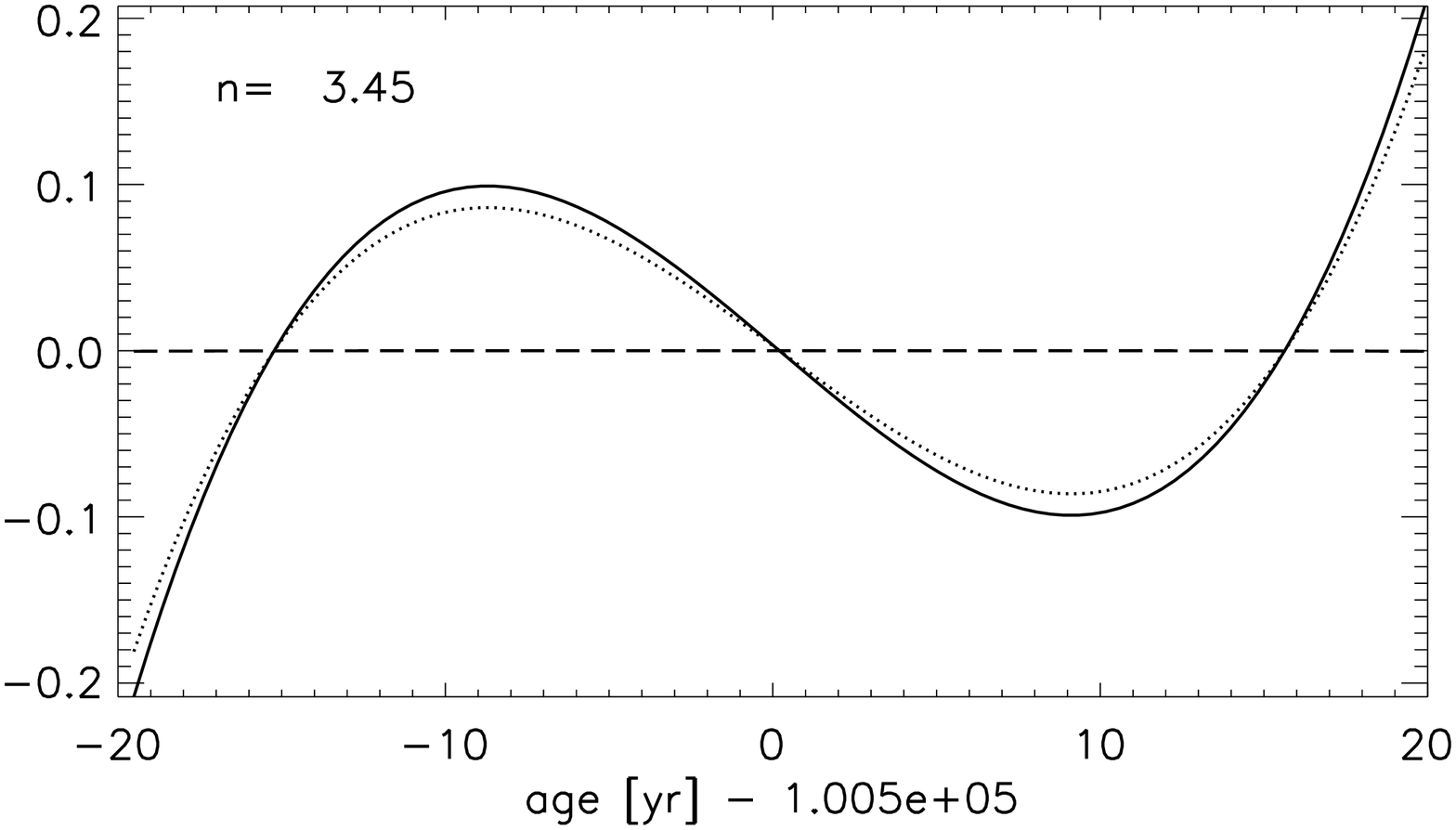}
\caption{Phase residuals for model C around $t_0=10.93$ kyr (upper panel) and $t_0=100.5$ kyr (lower panel), 
with the corresponding value of $n$ indicated in each case.
We show residuals after removing the best-fit solution including quadratic terms (solid) or including cubic terms (long dashes); dotted lines show the cubic residuals obtained  assuming a constant $B_0$.}
\label{fig4}
\end{figure}

\subsection{Timing residuals.}
In timing analysis of radio pulsars, an important piece of information is the study of the residuals, i.e, the phase differences between the observed signal and the best-fit model, including the frequency and frequency derivative. As shown in detail in \cite{Hobbs10}, there is a rich variety of shapes in the residuals of radiopulsars.
As we have shown, the strong Hall-induced interplay between different multipoles and the toroidal field produces a complex evolution of $B_0$. To compare with observational timing analysis, we 
proceeded as follows: first, we choose a short interval of 40 yr in our simulations, centered on a fixed time denoted 
by $t_0$. This interval is similar to the longest periods for
which phase-coherent timing analysis for radio pulsars can be performed. 
Then, from our theoretical $B_0(t)$, we obtain $\nu(t)$ 
by integration of the classical spindown formula (Eq.~\ref{equ:spin-down-B(t)}) assuming an orthogonal rotator, 
and the phase $\Phi(t)=\int_{t_0}^{t} \nu(t')dt'$. Finally, we fit our
synthetic time-dependent phase with a quadratic function
\begin{equation}
\Phi(t) = \Phi_0 + \nu_f (t-t_0) + \frac{1}{2}\dot{\nu}_f (t-t_0)^2 ~.
\end{equation}
The results of the fit, $\nu_f$ and $\dot{\nu}_f$, are of course in the range of values
of the {\it real} quantities in the time interval.

In Fig.~\ref{fig4} we show the phase residuals for model C (solid lines) at $t_0=10.93$ kyr (top) and $t_0=100.5$ kyr (bottom). 
During the two analyzed intervals, the mean values of the dipolar magnetic field are $\bar{B_0}=8.1\times 10^{12}$ G and $\bar{B_0}=5.3\times 10^{12}$ G. The cubic shape
indicates that the residuals are dominated by the next term in the Taylor series (i.e., red noise). Including in the 
fitting function a cubic term allows one to measure $\ddot{\nu}$ and the braking index ($n=0.39$ and $n=3.45$, 
respectively). With dashes we show the fourth-order residuals after subtracting the third-order term in the fitting function. 
For comparison, we also show with dotted lines the third-order residuals obtained assuming a constant value $B_0(t)=\bar{B_0}$ (that leads to $n=3$). 

In the first time interval, the change of magnetic field, $\delta B_0=1.3\times 10^{10}$ G $=1.6\times 10^{-3} \bar{B_0}$, 
is strong enough for the cubic residuals to be visibly different from the constant field case. The value of $|\ddot{\nu}|$ 
(and of $|n|$) is low, and the residuals of fourth order are significant. In the second time interval, $\delta B_0=1.4\times 10^8$ G $=2.6\times 10^{-5} \bar{B_0}$, and $n$ is close to 3. As a consequence, the deviation from the constant field spin-down behavior is slow and fourth-order residuals are orders of magnitude weaker (dashed line).

\begin{figure}[t]
\centering
\includegraphics[width=0.42\textwidth]{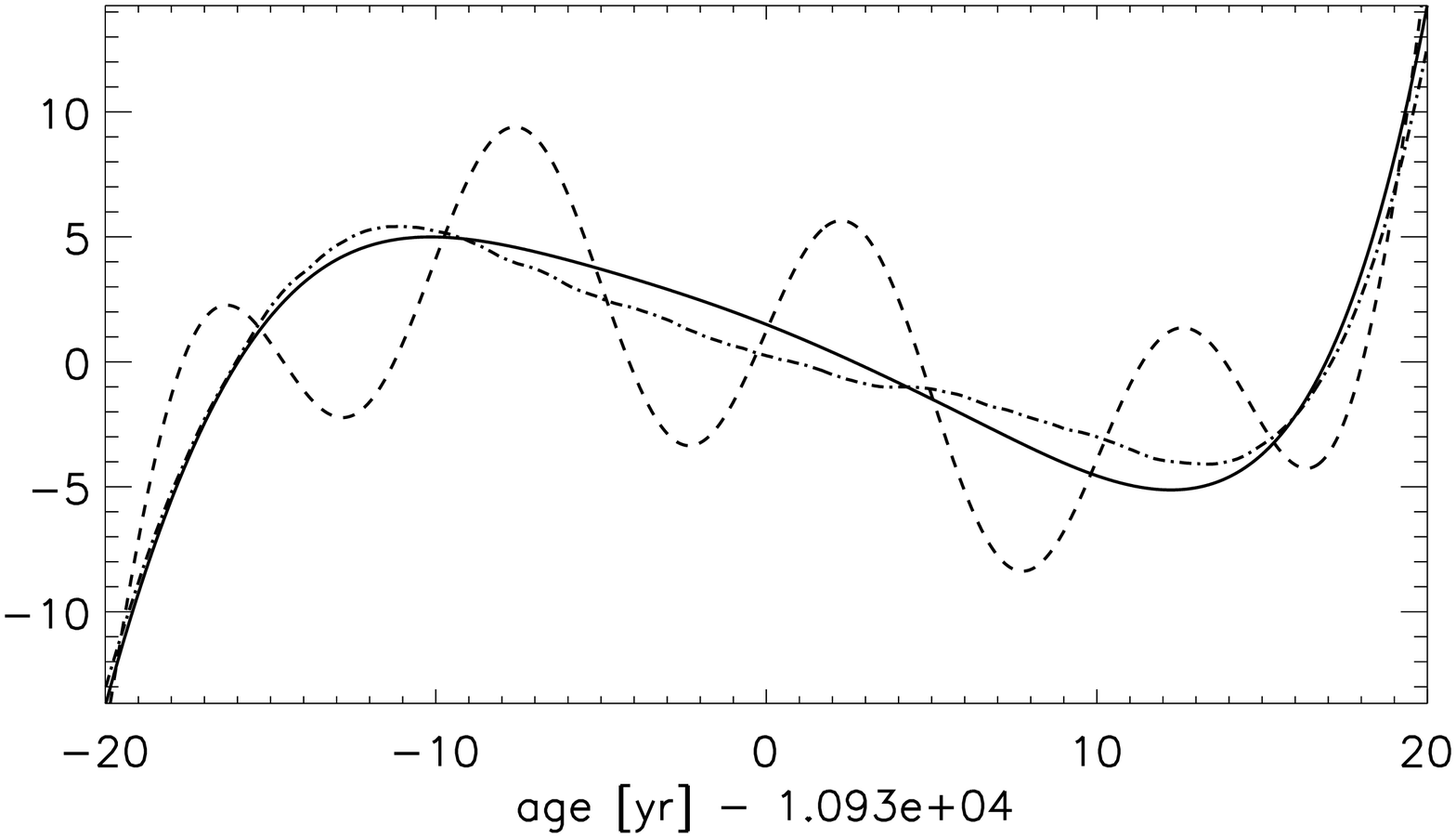}
\caption{Comparison of the cubic residuals of the theoretical model (solid line) with the result after adding
a sinusoidal perturbation with frequency $0.1$ yr$^{-1}$ (dashed) or a random perturbation (dot-dashed).
For both perturbation forms we considered a maximum amplitude of $(\delta B_0/2)=6.5\times 10^9$ G.
}
\label{fig5}
\end{figure}

To investigate the effect of short-time irregularities, we repeated the process but 
artificially added two additional sources of ``noise"  ($B_p$) to our theoretical values of $B_0(t)$,
with amplitude of the same order as the theoretical variations $\delta B_0$. 
In Fig.~\ref{fig5} we show the residuals obtained by adding a sinusoidal perturbation, 
$B_p=(\delta B_0/2)\sin(2\pi t/T)$, with $T=10$ yr (dashed lines), or a 
random perturbation of maximum absolute value $\delta B_0/2$ (dash-dotted lines). 
In our analysis, we sampled values every $\sim 10^{-2}$ yr, comparable with the typical integration time in observational data.
We checked that increasing the time interval strongly reduces
the contribution of the random noise at a fixed amplitude, as expected. Similarly, the periodic short-term noise
is reduced when the integration interval becomes much longer than the perturbation period.

Note that small perturbations of $B_0$ produce visible effects,
but the cubic residual still dominates. In constrast, in many pulsars a smooth cubic phase residual is clearly seen, 
which means that the spin-down is very stable, short-term irregularities are completely negligible (or averaged out in
the long observation period), and the measure of a braking index (even a high value) dominated by the 
secular evolution is robust. Glitches should appear as cusps in these plots (e.g. see data for PSR B0154+61
in \cite{Hobbs10}), but a more detailed investigation in this direction is beyond the scope of this paper.

\section{Conclusions.}

According to our current understanding of the magneto-thermal evolution of neutron stars, the time variation of the
magnetic field notably affects the braking index of pulsars and
contributes to the so-called red noise in the timing residuals. The qualitative picture seems to agree with the data:
  i) there is a short stage in which the dipolar field appears to be increasing, possibly reflecting the reemergence of magnetic field after initial accretion;
 ii) a stage dominated by Ohmic dissipation, while the star is still warm, and during which we always expect $n>3$; 
iii) once the star has cooled down, after about $10^5$ yr, oscillatory Hall-drift modes in the crust are excited, and quasi-periodic oscillations of the braking index are consequently expected.

This is by no means the only contribution to the timing noise, nor necessarily the dominant one.
For example, changes in the effective moment of inertia as the superconducting region
of the core grows and unknown mechanisms for alignment of the rotation and magnetic field
\citep{TM1998,Young2010} would have similar long-term effects.
In addition to this secular evolution, short-term magnetospheric effects, such as oscillations of the inclination angle around the equilibrium position, or magnetospheric changes \citep{Spit2012L,Spit2012}, could be also responsible for the strong quasi-periodic features seen in phase residuals for several cases \citep{Lyne2010}.

Our results show that realistic magnetic field evolution models predict temporal 
variations consistent with the observed trends. In particular, configurations similar to our model C could explain the
wide range in positive and negative values that are observed. This model has a moderate dipolar component
($10^{13}$ G at birth, which becomes a few times $10^{12}$ G at middle age), typical of pulsars, but a strong
octupolar component, large enough to activate Hall modes. The same generic behavior is expected for other models
with strong, smaller scale components, even if the dipole is weak.
However,  it must be mentioned that the models we studied
predict high values of the braking index ($10$-$100$) but not extremely high values ($10^3$-$10^4$), as reported
for pulsars with old characteristic ages. The absolute value of the braking index at late times, which can be
of either sign, is not simple to predict, since it depends on particular details of important components of the
magnetic field (higher order multipoles, toroidal component). Perhaps different field geometries 
(e.g., without axial symmetry) and/or magnetospheric corrections may account for the discrepancy.

We emphasize that there is absolutely no reason to expect that the magnetic field remains constant during
a pulsar lifetime, and we must abandon the oversimplified models that assume constant (in time) pure dipolar fields.
On the other hand, since there is a strong interplay between the magnetic, thermal, and rotational evolution 
of neutron stars, there are unexpected ways to obtain information. If future observations (or a deeper analysis of
archival data) increase the number
of reliable measures of braking indices, and we can firmly establish at which age the transition from the purely diffusive to the
oscillatory regime happens, there is coded information about the cooling history of neutron stars of
potential interest to place constraints on the interior physics.
In this respect, it is also important to mention that precise phase-connected 
timing solutions can also be performed in other bands, as has been recently done for 17 $\gamma$-ray-selected pulsars
\citep{Ray2011}, using the Large Area Telescope (LAT) on the Fermi Gamma-ray Space Telescope. This and other
future high-energy missions can enlarge our data sample and improve the precision of our measurements.

\begin{acknowledgements}
This research was supported by the grants AYA 2010-21097-C03-02 and ACOMP/2012/135.
DV is supported by a fellowship from the \textit{Prometeo} program for research groups of excellence of the Generalitat Valenciana (Prometeo/2009/103). We thank C. Espinoza and J.A. Miralles for their valuable comments.
\end{acknowledgements}



\end{document}